\documentstyle{article}

\def\abstract#1{\vskip 7mm
        \begin{center}{\large Abstract}\par \smallskip
                \begin{minipage}[c]{12cm}
                        \small #1
                \end{minipage}
        \end{center}
}
\def\title#1{\begin{center}{\Large\bf #1}\end{center}}
\def\author#1{\vskip 5mm \begin{center}{#1}\end{center}}
\def\address#1{\begin{center}{\it #1}\end{center}}

   \newcommand{\R}{ {\bf R} }
   \newcommand{\beq}[1]{\begin{equation}\label{#1}}
   \newcommand{\eeq}{\end{equation}}
   \newcommand{\bear}[1]{\begin{eqnarray}\label{#1}}
   \newcommand{\ear}{\end{eqnarray}}

\makeatletter
\@ifundefined{lesssim}{\def\lesssim{\mathrel{\mathpalette\vereq<}}}{}
\@ifundefined{gtrsim}{}{}
\def\vereq#1#2{\lower3pt\vbox{\baselineskip1.5pt \lineskip1.5pt
\ialign{$\m@th#1\hfill##\hfil$\crcr#2\crcr\sim\crcr}}}
\makeatother

\begin{document}

\title{%
  Problems of G and Multidimensional Models
}
\author{%
Vitaly N. Melnikov\footnote{E-mail: melnikov@rgs.phys.msu.su},
Vladimir D. Ivashchuk\footnote{E-mail: ivas@rgs.phys.msu.su}
}

\address{%
Center for Gravitation and Fundamental Metrology, \\
VNIIMS, 3-1 M. Ulyanovoy Str., Moscow, 117313, Russia \\
and \\
Institute of Gravitation and Cosmology,  \\
Peoples' Friendship University of Russia, \\
6 Miklukho-Maklaya St., Moscow 117198, Russia
}

\abstract{
 The relations for G-dot in multidimensional model
 with Ricci-flat internal space and multicomponent
 perfect fluid are obtained. A two-component example:
 dust + 5-brane, is also considered.
}

\section{Introduction}

 Dirac's Large Numbers Hypothesis (LNH) is the origin of many theoretical
 explorations of time-varying G.  According to LNH, the value of
 $\dot{G}/G$  should be approximately  the Hubble rate.
 Although it has become clear in recent
 decades that the Hubble rate is too high to be compatible with experiment,
 the enduring legacy of Dirac's bold stroke is the acceptance by modern
 theories of non-zero values of $\dot{G}/G$ as being potentially consistent
 with physical reality.
 There are three problems related to $G$, which origin lies mainly in
 unified models predictions:
1) absolute $G$ measurements,
2) possible time variations of $G$,
3) possible range variations of $G$ -- non-Newtonian, or new interactions.
For 1) and 3) see \cite{Mel}.

After the original {\it Dirac hypothesis} some new ones appeared
and also some generalized {\em theories} of gravitation admitting the
variations of an effective gravitational coupling. We may single out three
stages in the development of this field:
\begin{enumerate}
\item Study of theories and hypotheses with variations of FPC,
 their predictions and confrontation with experiments (1937-1977).

\item
Creation of theories admitting variations of an effective gravitational
constant in a particular system of units, analyses of experimental and
observational data within these theories \cite{Stan}
(1977-present).

\item
Analyses of FPC variations within unified models \cite{Mel}
(present).
\end{enumerate}

Different theoretical schemes lead to temporal variations of the
effective gravitational constant:
\begin{enumerate}
\item
Empirical models and theories of Dirac type, where $G$ is replaced by
$G(t)$.

\item
Numerous scalar-tensor theories of Jordan-Brans-Dicke type where $G$
depending on the scalar field $\sigma (t)$ appears.

\item
Gravitational theories with a conformal scalar field arising in different
approaches \cite{Stan}.

\item
Multidimensional unified theories in which there are dilaton fields and
effective scalar fields appearing in our 4-dimensional spacetime from
additional dimensions \cite{Mel}. They may help also in solving the
problem of a variable cosmological constant from Planckian to present
values.
\end{enumerate}

 A striking feature of most modern
 scalar-tensor and  unification theories, is that they do not admit a
 unique and universal constant values of physical constants and
 of the Newtonian gravitational coupling  constant $G$ in particular.
 In this paper we briefly set out the results of some
 calculations which have been carried out for various theories, and we
 discuss various bounds  that may be suggested
 by multidimensional theories. Although the bounds on G-dot and $G(r)$
 are in some classes of theories rather wide on purely theoretical grounds
 as a result of adjustable parameters, we note that observational data
 concerning other phenomena may place limits on the possible range of
 these adjustable parameters.

 Here we limit ourselves to the problem of G-dot (for $G(r)$
 see \cite{Stan,4,5,Mel}).  We show that the various theories predict the
 value of $\dot{G}/G$ to be $10^{-12}/yr$ or less.  The significance of this
 fact for experimental and observational determinations of the value of or
  upper bound on G-dot is the following:  any determination with error
 bounds significantly below $10^{-12}$  will typically be compatible with
 only a small portion of existing theoretical models and will therefore cast
 serious doubt on the viability of all other models.  In short, a tight
 bound on G-dot, in conjunction with other astrophysical observations, will
 be a very effective "theory killer."

 Some estimations for G-dot were done long ago in the frames of
 general scalar tensor theories using values of cosmological parameters
 ($\Omega$, $H$, $q$ etc) \cite{Stan,Mel}.
  With modern values they predict $\dot{G}/G$ at the level of
  $10^{-12}/yr$ and less (see recent estimations of A. Miyazaki \cite{Mi},
  predicting time variations of $G$ at the level of $10^{-13} yr^{-1}$)
  for the Machian-type cosmological solution in the Brans-Dicke theory).

  The most reliable experimental bounds on $\dot{G}/G$
  (radar ranging of spacecraft dynamics \cite{Hel})
  and laser lunar ranging \cite{Dic} give the limit
  of $10^{-12}/yr$).

\section{G-dot in $(4+N)$-dimensional cosmology
with multicomponent anisotropic fluid }

We consider here a $(4+N)$-dimensional cosmology
with an isotropic 3-space and an arbitrary Ricci-flat internal space. The
Einstein equations provide a relation between ${\dot{G}/G}$ and other
cosmological parameters.

\subsection{The model}

Let us consider $(4+N)$-dimensional theory
described by the action
\begin{equation}
S_g = \frac{1}{2\kappa^2}\int d^{4+N}x\sqrt{-g}R\ ,
\end{equation}
where $\kappa^2$ is the fundamental gravitational constant. Then the
gravitational field equations are
\begin{equation}
R^M_P = \kappa^2(T^M_P-\delta^M_P\frac{T}{N+2})\ ,
\end{equation}
where $T^M_P$ is a $(4+N)$-dimensional energy-momentum tensor,
$T=T^M_M$, and  $M,P=0,...,N+3$.

For the $(4+N)$-dimensional manifold we assume the structure
\begin{equation}
M^{4+N} = R\times M^3_k\times K^N\ ,
\end{equation}
where $M^3_k$ is a 3-dimensional space of constant curvature, \\
$M^3_k=S^3,\
R^3,\ L^3$ for $k=+1,0,-1$, respectively, and $K^N$ is a $N$-dimensional
compact Ricci-flat Riemann manifold.

The metric is taken in the form
\begin{equation}
g_{MN}dx^Mdx^N = - dt^2 + a^2(t)g^{(3)}_{ij}(x^k)dx^idx^j +
b^2(t)g^{(N)}_{mn}(y^p)dy^mdy^n \ ,
\end{equation}
where $i,j,k=1,2,3;\ m,n,p=4,...,N+3;\ g^{(3)}_{ij},\ g^{(N)}_{mn},\ a(t)$
and $b(t)$ are, respectively, the metrics and scale factors for $M^3_k$
and $K^N$.  For $T^M_P$ we adopt the expression
of the multicomponent (anisotropic) fluid form
\begin{equation}
(T^M_P) = \sum_{\alpha = 1}^m
diag(- \rho^{\alpha} (t),  p_3^{\alpha}(t)\delta^i_j,
p_N^{\alpha}(t)\delta^m_n) .
\end{equation}

Under these assumptions the Einstein equations take the form
\bear{6}
\frac{3\ddot{a}}{a} + \frac{N\ddot{b}}{b} =
\frac{\kappa^2}{N+2} \sum_{\alpha = 1}^m
[-(N+1)\rho^{\alpha}
-3p_3^{\alpha} -Np_N^{\alpha}], \\
\label{7}
\frac{2k}{a^2} + \frac{\ddot{a}}{a} + \frac{2\stackrel{.}{a}^2}{a^2} +
\frac{N\stackrel{.}{a}\stackrel{.}{b}}{ab} =
\frac{\kappa^2}{N+2}  \sum_{\alpha = 1}^m
[\rho^{\alpha} +
(N-1)p_3^{\alpha} - Np_N^{\alpha}], \\
\label{8}
\frac{\ddot{b}}{b} + (N-1)\frac{\stackrel{.}{b}^2}{b^2} +
\frac{3\stackrel{.}{a}\stackrel{.}{b}}{ab} =
\frac{\kappa^2}{N+2}  \sum_{\alpha = 1}^m
[\rho^{\alpha} -3p_3^{\alpha}+ 2p_N^{\alpha}].
\ear

The 4-dimensional density is
\beq{9}
\rho^{\alpha,(4)}(t) = \int_K d^Ny\sqrt{g^{(N)}}b^N(t)\rho^{\alpha}(t) =
\rho^{\alpha}(t)b^(t),
\eeq
where we have normalized the factor $b(t)$ by putting
\beq{10}
\int_K d^Ny\sqrt{g^{(N)}} = 1.
\eeq
On the other hand, to get the 4-dimensional gravity equations one should put
$8\pi G(t)\rho^{\alpha(4)}(t) = \kappa^2\rho^{\alpha}(t)$.
Consequently, the effective
4-dimensional gravitational ``constant'' $G(t)$ is defined by
\begin{equation}
8\pi G(t) = \kappa^2b^{-N}(t)
\end{equation}
whence its time variation is expressed as
\begin{equation}
\dot{G}/G = -N \dot{b}/b.
\end{equation}

\subsection{Cosmological parameters}

Some inferences concerning the observational cosmological parameters can be
extracted just from the equations without solving them
\cite{BIM1}. Indeed, let us define
the Hubble parameter $H$, the density parameters $\Omega^{\alpha}$ and
the "deceleration"
parameter $q$ referring to a fixed instant $t_0$ in the usual way
\begin{equation}
H = \dot{a}/a,  \qquad
\Omega^{\alpha} = 8\pi G\rho^{\alpha,(4)}/3H^2 =
\kappa^2\rho^{\alpha}/3H^2, \quad  q = -a  \ddot{a}/\dot{a}^2\ .
\end{equation}
Besides, instead of $G$ let us introduce the dimensionless parameter
\begin{equation}
g = \dot{G}/GH = -Na \dot{b}/\dot{a}b .
\end{equation}

Then, excluding $b$ from (\ref{6}) and (\ref{8}), we get
\beq{14}
\frac{N-1}{3N} g^2 -g +
q - \sum_{\alpha = 1}^m A^{\alpha} \Omega^{\alpha} = 0
\eeq
with
\beq{15}
A^{\alpha} = \frac{1}{N+2} [2N+1+3(1-N)\nu_3^{\alpha} +
3N \nu_N^{\alpha}],
\eeq
where
\beq{16}
\nu_3^{\alpha} = p_3^{\alpha}/\rho^{\alpha}, \quad
\nu_N^{\alpha} = p_N^{\alpha}/\rho^{\alpha}, \quad   \rho^{\alpha} >0\ .
\eeq

When $g$ is small we get from (\ref{14})
\beq{17}
g \approx q  -  \sum_{\alpha = 1}^m A^{\alpha} \Omega^{\alpha}.
\eeq

Note that (\ref{17}) for $N=6$, $m=1$,
$\nu_3^{1}=\nu_6^{1}=0$ (so that $A^{1}=13/8$)
coincides with the corresponding
relation of Wu and Wang \cite{WW}
obtained for large times in case $k=-1$
(see also \cite{IM1}).

If $k=0$, then in addition to (\ref{17}),
one can obtain a separate relation
between $g$ and $\Omega^{\alpha}$, namely,
\beq{18}
\frac{N-1}{6N} g^2 - g + 1 - \sum_{\alpha = 1}^m \Omega^{\alpha} = 0
\eeq
(this follows from the Einstein equation
$R^0_0-\frac{1}{2}R= \kappa^2T^0_0$, which is certainly
a linear combination of (\ref{6})-(\ref{8}).

The present observational upper bound on $g$ is
\beq{18a}
\mid g \mid \lesssim 0.1
\eeq
if we take in accord with  \cite{Hel,Dic}
\beq{18b}
\mid \stackrel{.}{G}/G \mid \lesssim
0.6 \times 10^{-11}(y^{-1})
\eeq
and $H = (0.7 \pm 0.1) \times 10^{-11}(y^{-1})
\approx 70 \pm 10 (km/s.Mpc)$.

\subsection{Two-component example: dust + $(N-1)$-brane}

Let us consider two component case: $m= 2$. Let
the first component (called "matter") be a dust, i.e.
\beq{19}
\nu_3^{1}  = \nu_N^{1} = 0,
\eeq
and the second one (called "quintessence") be a $(N-1)$-brane,
i.e.
\beq{20}
\nu_3^{2}  =  1, \qquad  \nu_N^{2} = - 1.
\eeq

We remind   that as it was mentioned in \cite{IMbil}
the multidimensional cosmological model
on product manifold $\R \times M_1 \times ... \times M_n$
with fields of forms
(for review see \cite{IMtop})
may be described in terms of multicomponent
"perfect" fluid \cite{IMfl} with the following equations
of state for $\alpha$-s component:
$p_i^{\alpha} =  - \rho^{\alpha}$ if $p$-brane worldvolume
contains $M_i$ and
$p_i^{\alpha} =   \rho^{\alpha}$ in opposite case.
Thus, the field of form matter leads us either to
$\Lambda$-term, or to stiff matter equations of state
in internal spaces.

In this case we get from (\ref{17}) for small $g$
\beq{21}
g \approx q  -  \frac{2N+ 1}{N+2} \Omega^{1} +
4 \frac{N - 1}{N+2} \Omega^{2},
\eeq
and for $k =0$ and small $g$  we obtain from (\ref{18})
\begin{equation}
1 -g  \approx \Omega^{1} + \Omega^{2}.
\end{equation}

Now we illustrate the formulas by the following example
when  $N =6$ ($K^6$ may be a Calabi-Yau manifold) and
\beq{22}
-q  =  \Omega^{1} = \Omega^{2} = 0.5.
\eeq
We get from (\ref{21})
\beq{23}
g \approx - \frac{1}{16} \approx -0.06
\eeq
in agreement with (\ref{18a}).

In this case the second fluid component corresponds to
magnetic (Euclidean) $NS5$-brane (in $D=10$ type I, Het or II A
string models).

\end{document}